\documentstyle[revtex]{aps}
\begin{document}
\draft
\hspace{5 in} {\bf Preprint IFUM 429/FT}

\hspace{5.3 in} {\bf July 1992}
\vspace{15 mm}
\begin{title}
A quantitative study of the Kosterlitz-Thouless phase\\
transition in a system of two-dimensional plane rotators\\
( XY model ) by high temperature expansions through $\beta^{20}$
\end{title}
\author{P. Butera and M. Comi}
\begin{instit}
Istituto Nazionale di Fisica Nucleare\\
Dipartimento di Fisica, Universit\`a di Milano\\
Via Celoria 16, 20133 Milano, Italy
\end{instit}
\begin{abstract}
High temperature series
expansions of the spin-spin correlation
function for the plane rotator (or XY) model
on the square lattice are  extended
by three terms through order
$\beta^{20}$. Tables of the expansion coefficients
are reported for the correlation function
spherical moments of order $l=0,1,2$.
The  expansion coefficients
through $\beta^{15}$ for the vorticity are also tabulated.
Our analysis of the
series supports the
Kosterlitz-Thouless predictions on
the structure of the critical
singularities and leads to fairly
accurate estimates of the
critical parameters.

\end{abstract}
\section{ Introduction }

The critical  behavior of the
two-dimensional plane rotator (or XY) model,
 has long been studied numerically
by  high
temperature expansions (HTEs)[\cite{htstudies,gut}],
by hamiltonian strong coupling expansions[\cite{hamer}],
by MonteCarlo  (MC) simulations[\cite{oldmc,seiler}],
and by other techniques.
In spite of these considerable efforts,
an accurate verification of
the Kosterlitz and Thouless (KT) theory [\cite{kt,bere}]
remained out of reach
before recent technical advances such as
the calculation of  long high
temperature expansions (HTEs)[\cite{bcm}],
  the recent invention of   MC algorithms
with reduced critical slowing-down [\cite{sokal}] and
the availability
of a greater computing power.
In the last few years  many  extensive numerical studies
of an increasing accuracy
have appeared[\cite{bcm}], [\cite{gupta,wolff,edwards}],
 [\cite{janke}], [\cite{biferale}],
 part of which have  been also
stimulated  by a challenge
to the KT approach issued in Ref.[\cite{seiler}].
 These works generally
 favor the essential singularity structure predicted
by KT arguments over the
power-law  structure of usual critical phenomena.
 However, one more
 warning for caution on the actual
scope and limits of  MC results
came from Ref.[\cite{edwards}], reporting
an exemplary analysis of a multigrid
MC simulation.
 This study sets higher qualitative  standards for
  future  MC work
and, like Ref.[\cite{seiler}], again questions the
possibility of discriminating between the KT and the
 power-law scenarios, only by  fits
to MC data with the present level of accuracy and extension.
On the other hand, we had stressed [\cite{bcm}]  that,
even in the absence of a   detailed  rigorous
theoretical treatment of the
XY model, the general attitude in favor of the KT
picture can be convincingly justified,
already now, if
all available numerical evidence both
from the simulations and from
the newly computed HTEs
is properly taken into account.

Here we present  a  further extension (by three terms
 up to order $\beta^{20}$)
and a new analysis of HTE for the 2D
plane rotator model on the
square lattice.
A HTE approach is always a necessary complement to
the statistical simulations since it provides detailed
and extensive information,
but in this case it  also improves  significantly our
chances to distinguish numerically between the KT and
the power-law behavior and leads us to
exclude this latter possibility.
Once the question concerning the
nature of the critical singularity
is settled,
we can get reliable  estimates
of the critical parameters, although,
in this case, perhaps
less precise
than it  could be expected
from series of such a length.

The paper is organized as follows: in
the second Section the definitions
of the quantities that have been computed
are briefly recalled and
 their HTE coefficients are tabulated.
The third Section is devoted
to an analysis of the series by ratio-extrapolation,
and by rational and differential approximants techniques.
The last Section contains some discussion
of previous work
 and our conclusions.

\section{ The HT series}
\label{sec:ht}

The Hamiltonian  of the two-dimensional
plane rotator (or XY) model is
\begin{equation}
H\{s\}= -\sum_x \sum_{ \mu=1{,}2 } s(x) \cdot s(x+e_\mu).
\label{eq:hamilt} \end{equation}
Here $s(x)$ is a two-component classical spin
of unit length associated
to the site  with position vector

$ x=n_1 e_1+n_2 e_2 =(n_1,n_2)$ of a 2-dimensional square
lattice
 and $ e_1$, $ e_2$ are the two elementary lattice vectors.
The sum over $ x $
extends over all lattice sites.

Our series  have been computed by  a FORTRAN code
which solves iteratively the Schwinger-Dyson
equations for the correlation
functions[\cite{bcm,guerra,march}].
The algorithm has been described
in full detail in  Ref.[\cite{bcm}].
Here  it is enough  to mention that we have computed
the HTE coefficients of the
two-point correlation function
\begin{equation}
 C(x;\beta)=<s(0)\cdot s(x)>
\label{eq:corr} \end{equation}
for the 120 inequivalent sites $x$
for which the expansion is
 non trivial to order $\beta^{20}$.
In this approach the main obstacle to a further
extension of our results is not computational
time which is still definitely modest
( of the order of 20 hours for a 3500 VAX station),
but the increasing demand of fast memory.
Our work has been made possible by a
 laborious segmentation
of the computing procedure.

The object of our analysis
are the series for the spherical moments of
the correlation function  $m^{(l)}(\beta)$
 defined as follows:
\begin{equation}
 m^{(l)}(\beta)=\sum_x \vert x\vert^{l} C(x;\beta)
= \sum_{r=1}^\infty
 a^{(l)}_r \beta^r ,
\label{eq:moment} \end{equation}
(here $\vert x \vert =\sqrt{n_{1}^{2}+n_{2}^{2}}$),
$l \geq 0 $ and the sum extends over all lattice sites.
The zero-th order spherical moment  $m^{(0)}(\beta)$ is
 also called (reduced)
susceptibility and denoted by $\chi(\beta)$.
The data we are presenting augment
significantly our earlier work [\cite{bcm}].

In Table I,II and III we have reported the HTE
coefficients through $\beta^{20}$
of the spin-spin correlation functions
$<s(0)\cdot s(x)>$ with
$x=(1,0) , x=(2,0)$ and $ x=(1,1)$, respectively.

In Tables IV,V and VI we have reported the
expansion coefficients for the moments
$ m^{(l)}(\beta) $
with $ l= 0$, $1$, $2$.

In Table VII we  have reported
the HTE coefficients through
$\beta^{15}$ of the expectation value of the squared
vorticity $v(\beta)^2$,
a quantity built in terms of two-,three-
and four-spin correlation
functions [\cite{onofri}] which probes the
vortex pair dissociation mechanism of
the phase transition.
In the definition of the vorticity
it is convenient to refer to
the representation $ s(x) = (cos\theta(x), sin\theta(x)) $,
and then we have
\begin{eqnarray}
<v(\beta)^2> = {1 \over 3}+{4 \over \pi^{2}}
\sum_{n=1}^\infty {(-1)^{n}
\over n^{2}}<e^{in(\theta_1-\theta_2)}>
-{2 \over \pi^{2}}\sum_{n,m \not=0}{(-1)^{n+m}
\over nm}<e^{in(\theta_1-\theta_2)+
im(\theta_2-\theta_3)}>\nonumber\\
-{1 \over \pi^{2}}\sum_{n,m \not=0}
{(-1)^{n+m} \over nm}<e^{in(\theta_1-\theta_2)+
im(\theta_3-\theta_4)}>. \end{eqnarray}
Here $ \theta_{1}, \theta_{2}, \theta_{3}, \theta_{4} $ are
the angular variables associated to the four sites defining
an elementary square on the lattice.
We shall only tabulate this series,
since it has already been
extensively discussed in Ref.[\cite{onofri}],
where our HTE has been
compared to a Langevin simulation.

Finally, we remind the interested reader that a  list of the
presently available HT data for this model also includes a
series [\cite{bcm}] through
$\beta^{12}$ for the "true correlation length"
[\cite{fisher}]
and a
series through $\beta^{14}$ for $\chi^{(4)}(\beta)$,
the second derivative of the susceptibility
with respect to the magnetic
field at zero field [\cite{luscher}].

\section{ An analysis of the HT series }
\label{sec:analys}
 In this Section we present the
estimates of the critical parameters obtained by
simple methods of series analysis [\cite{bcm,seran,nickel,diffapp}]
which, after
some numerical experimenting
both with appropriate model series and with our series,
turned out to be
best suited for extracting the expected behavior
of the correlation moments in the critical region.

Let us first recall briefly the main results of
the non-rigorous renormalization
group analysis of the plane rotator model [\cite{kt,bere}].

The correlation length $\xi(\beta)$
is expected to diverge as
$ \beta \uparrow \beta_{c}$
with the unusual singularity
\begin{equation}
\xi(\beta) \propto \xi_{as}(\beta)=  exp(\frac {b}
{\tau^{\sigma}})[ 1 + O(\tau)]
\label{eq:corleng} \end{equation}
where $\tau=\beta_{c}-\beta$.

The value of the exponent $ \sigma$
predicted in Ref.[\cite{kt}]  is
 $ \sigma=1/2 $ and $b$ is a non-universal
positive constant.

At the critical temperature, the
asymptotic behavior of the two-spin
correlation function as
$ r=|x| \rightarrow \infty$ is expected to be
\begin{equation}
<s(0)\cdot s(x)> \propto
\frac{({\rm ln}(r))^{2\theta}} {r^{\eta}}
[1+O({\rm ln}({\rm ln}(r))/{\rm ln}(r))]
\label{eq:asycor} \end{equation}
The values predicted [\cite{kt,amit}]  for $ \eta$   and
$ \theta$ are, respectively, $ \eta=1/4 $ ,
$ \theta=1/16 $.

{}From  Eqs. (\ref{eq:corleng})
and (\ref{eq:asycor}) it follows that,
for $ l > \eta-2 $,  the
 correlation moment $m^{(l)}(\beta)$ should
diverge as
$ \beta \uparrow \beta_{c}$ with the singularity

\begin{eqnarray}
m^{(l)}(\beta) \propto
\tau^{-\theta}\xi_{as}(\beta)^{2-\eta+l}
[ 1 + O(\tau^{\frac {1} {2}}
{\rm ln}(\tau))]
\label{eq:asymom} \end{eqnarray}

At $\beta_{c}$  a line of critical
points should begin which extends to $ \beta=\infty$,
so that for $ \beta >\beta_{c}$
both $ \xi$ and the correlation moments remain infinite.

Finally, we must recall that the existence
of a transition of the system
from a vortex-dominated high temperature phase
to a spin-wave low temperature phase
 has been  proved [\cite{frohl}]
and that the lower bound:
$\beta_{c}\geq {\rm ln}(1+\sqrt{2}) \approx 0.88$
has been established [\cite{aizen}]
for the critical inverse temperature.

In Ref. [\cite{bcm}] we  used a
theorem of  Darboux [\cite{olver}]
to  point out  that  the
leading asymptotic behavior for large order
of the HTE  coefficients of $ \xi$
(and of the correlation moments)
may  be  estimated  by   saddle  point  approximation
of a contour integral, if it
is determined by the singularity (\ref{eq:corleng}).
Consider, for example,
 $ \xi(\beta)=\sum_n c_{n} \beta^{n}$,
then, for large $n$, we have
\begin{equation}
c_{n}=\frac {1} {2 \pi i}
\oint\xi(\beta)\frac {d\beta} {\beta^{n+1}}
\propto  \frac {1} {2 \pi i} \oint\xi_{as}(\beta)
\frac {d\beta} {\beta^{n+1}}
\label{eq:asycoeff} \end{equation}
For general $ b,\sigma>0 $ the following
 asymptotic expression is  obtained [\cite{bcm}]:

\begin{equation}
c_{n} \propto \beta_{c}^{-n-1} exp[B(n+1)^{\sigma \epsilon}
+ O(n^{(\sigma-1)\epsilon})]
\label{eq:asycoeffc} \end{equation}
with
\begin{equation}
\epsilon=\frac {1} {1+\sigma} \qquad and \qquad
 B=\frac {(\sigma+1)b^{\epsilon}}
{(\sigma \beta_{c})^{\sigma \epsilon}}.
\label{eq:asypar} \end{equation}
Therefore for the ratios of the
successive HTE coefficients
$ r_{n}(\xi)=c_{n}/c_{n+1}$ we have
\begin{equation}
 r_{n}(\xi)=\beta_{c}+\frac {C}
{(n+1)^{\epsilon}}  +O(1/n^{\lambda})
\label{eq:ratiokt} \end{equation}
with $C=-(\sigma b
  \beta_{c})^{\epsilon}$
 and $\lambda = min(1,2\epsilon)$.
A similar formula is valid for the ratios
$r_n(m^{(l)})= a^{(l)}_n/a^{(l)}_{n+1}$
of the successive HTE coefficients of
 the  correlation moment
$ m^{(l)}(\beta)$ if $C$ is replaced
by $C_{l}=-((2-\eta+l) \sigma b  \beta_{c})^{\epsilon}$.
The correction terms
$ O(1/n^{\lambda}) $ in (\ref{eq:ratiokt})
 also account for the
subdominant singularities in (\ref{eq:corleng}).
 According to
the KT prediction we should have $ \epsilon=2/3 $. This
 is a neat
 signature of the KT singularity in the HTE approach.

 On the other hand if, instead of (\ref{eq:corleng})
and (\ref{eq:asymom}), we had
conventional power-law critical singularities so that,
 as $ \beta \uparrow \beta_{c}$,
\begin{equation}
m^{(l)}(\beta) \sim
\tau^{-\gamma -l\nu}
[ A_{l} + B_{l} \tau^{\Delta}+..]
\label{eq:asymompow} \end{equation}
where, $\Delta > 0$ and, as usual,
$\gamma$ and $\nu$ denote
the susceptibility and the
correlation length exponents respectively,
we would obtain a formula analogous
to eq.  (\ref{eq:ratiokt})
with $\epsilon=1$ and ${\lambda = 1+\Delta}$,
namely
\begin{equation}
 r_{n}(m^{(l)})=\beta_{c}+\frac {\beta_{c}(1-\gamma-l\nu)} {n}
+O(1/n^{ 1+\Delta})
\label{eq:ratiopow} \end{equation}
A  complication for the series analysis is due to
the occurrence of an antiferromagnetic singularity
at $-\beta_{c}$ which is typical of loose lattices.
It should however affect only the higher correction
terms  in (\ref{eq:ratiokt})
( or in (\ref{eq:ratiopow})),
 usually introducing oscillations in the ratio plots.
A simple prescription to reduce
this inconvenience in numerical extrapolations
consists in  studying
the ratios of alternate coefficients: for example
$\bar r_{n}(m^{(l)})=
\sqrt{a^{(l)}_{n-1}/a^{(l)}_{n+1}}$
 instead of the usual ratios
$r_{n}(m^{(l)})=a^{(l)}_{n}/a^{(l)}_{n+1}$.

In view of the above considerations our analysis
 should begin
 by trying to estimate
 $\epsilon$ or, equivalently, $\sigma$.

Let us first perform the simplest
tests on the ratio sequences.

In Fig.1 we have plotted vs  $1/n$ the
sequences of alternate ratios
$\bar r_{n}(m^{(l)})$ for $l=0,1,2$.
 These ratio plots exhibit a
sizable curvature and an increasing slope for large $n$.
If eq.(\ref{eq:ratiopow}) were an adequate
representation of the asymptotic
behavior of $\bar r_{n}(m^{(l)})$,
we should be able to suppress
the $O(1/n)$ terms in (\ref{eq:ratiopow}) by forming the
linearly extrapolated sequences
\begin{equation}
\bar r^{(1)}_{n}(m^{(l)})= n\bar r_{n}(m^{(l)})-
(n-1)\bar r_{n-1}(m^{(l)})
=\beta_{c} + O(1/n^{ 1+\Delta})
\label{eq:extrapolin} \end{equation}
which, for large $n$,  should approach
with vanishing slope their common
limit $\beta_{c}$.
 However this does not  happen, as it is  shown
in Fig.1, where we have also plotted the
extrapolated sequences $\bar r^{(1)}_{n}(m^{(l)})$
 versus $1/n$. The estimates of $\beta_{c}$ thus obtained
 are still rapidly increasing with order.

Still under the assumption of power-law
critical singularities,
we might also  compute a
sequence of (unbiased) estimates of
$\gamma+ l\nu$ by the formula

\begin{equation}
(\gamma+ l\nu)_{n}=
\frac {(n-1)^2\bar r_{n}(m^{(l)})-n(n-2)\bar r_{n-1}(m^{(l)})}
{n\bar r_{n-1}(m^{(l)})-(n-1)\bar r_{n}(m^{(l)})}
\label{eq:gamplusnu} \end{equation}

We have reported the sequences of
estimates so obtained for $\gamma$ and $\gamma +\nu$
versus $1/n$ in Fig.2.
If any conclusion at
all may be drawn from these standard computations is that,
 under the assumption of
power-law scaling, a ratio analysis
might  suggest  that $\beta_{c} > 1.06 $,
$\gamma > 2.9 $ and $\gamma+\nu > 4. $.
The simplest extrapolations would suggest
 $\beta_{c} > 1.09 $,
$\gamma > 3.4 $ and $\gamma+\nu > 4.8 $.
As we shall discuss later,
these estimates are inconsistent with the
significantly smaller estimates resulting from fits
of MC data to power law critical behavior.

Let us observe now that, if eq.(\ref{eq:ratiokt})
is valid instead of eq.(\ref{eq:ratiopow}), then by
reporting the $ \bar r_{n}(m^{(l)})$ sequences
versus $1/n^{2/3}$, we should obtain nicely straight plots.
Fig.3 appears to be a convincing
illustration of this statement.
The next obvious step of
suppressing the $O(1/n^{2/3})$ terms
in the sequences $ \bar r_{n}(m^{(l)})$
 by forming the (nonlinearly) extrapolated sequences

\begin{equation}
\bar s_{n}(m^{(l)})=
\frac{n^{2/3}\bar r_{n}
(m^{(l)})-(n-2)^{2/3}
\bar r_{n-2}(m^{(l)})} {n^{2/3}-(n-2)^{2/3}}
=\beta_{c}+ O(1/n)
\label{eq:extrapokt} \end{equation}
 does not result in  a sequence  regular enough to warrant a
further extrapolation in $1/n$ and
therefore a more precise estimate of $\beta_{c}$.
We have, however, computed also the linearly extrapolated
 sequence $s^{(1)}_{n}(m^{(l)})$ and reported the results
 in Fig.3. From these we can infer that $\beta_{c}=1.120 \pm0.005$.
 A possible  improvement of this procedure
should be based on Euler transformed
moment series, as we have discussed at length
in Ref. [\cite{bcm}], where
the results and the conclusions of this kind of analysis
on a shorter series can be found. Since this
procedure might be questioned [\cite{nickel}],
 we will not insist here on the details.

We can give a direct unbiased estimate of
$\epsilon $ in terms of ratios as
follows.

Introduce the quantity
\begin{equation}
t_{n}  =\frac {\bar r_{n}(\chi^{2})} {\bar r_{n}(\chi)}
\label{eq:ratioeps} \end{equation}
then, if the ratios
$\bar r_{n}(\chi)$ and $\bar r_{n}(\chi^{2})$
have the asymptotic behavior
 (\ref{eq:ratiokt}), the
sequence
\begin{equation}
\epsilon_{n}=n {\rm ln}( \frac {t_n -1} {t_{n+1}-1} )
\label{eq:epsi} \end{equation}
 will provide estimates
of  $\epsilon$.
Quantities $u_{n}$ and $v_{n}$
analogous to $t_{n}$ may be defined in
terms of the moments $m^{(1)}$ and $m^{(2)}$
and their squares,
and, via  eq.
(\ref{eq:epsi}), the corresponding sequences $\epsilon_{n}'$  and
$\epsilon_{n}''$ may be formed.
All these sequences  have been plotted vs. $ 1/n $  in Fig.4.
They are slightly irregular
so that it is not easy to get precise extrapolations
 to $n=\infty$.  The figure, however, clearly suggests
that the sequences have a common
limiting value  somewhere around 0.67.
This result definitely excludes a power-law singularity
( in that case, of course, the limiting value should be 1)
and compares very nicely with the KT
prediction  $ \epsilon=2/3 $.

 Other prescriptions to compute $\epsilon$, as well as to
obtain a first estimate of the
exponent $\theta$, involve Pad\`e
approximants (PAs).

Before entering into the PA
analysis, let us  recall that PAs are
known to converge well to  (locally)
 meromorphic analytic functions.
In order to take advantage of this property,
it would be convenient to work with suitable
functions of the given HT series
for which the critical singularity is a simple pole.
Due to the structure (\ref{eq:corleng}),  (\ref{eq:asymom})
this is not possible in general,
and, at best, one can form functions of the HT series
 having a simple pole  accompanied by  subdominant  confluent
 singularities.
As a consequence,  whenever high precision estimates are
pursued from expansions of limited length, one should keep in mind that
the convergence properties and the accuracy
of the PA estimates may be  different, not only for
different quantities, but also for
different functions of the
same quantity, according to the
nature and relative strength of the
 subdominant singularities.
For instance, estimates of $\beta_c$ or $\sigma$  obtained from
different moments (or different functions of the
same moment)   may
differ more than the
(statistical) quantity (see below)
we shall adopt as an estimate of the error.
In order to reduce this  "systematical" error
 in  the analysis, one should
 resort  to  differential approximants (DAs), a
natural generalization[\cite{diffapp}] of the rational approximants
 which, unlike PAs, can make allowance numerically
for the confluent subdominant singularities.
Occasionally, we have also computed DAs,
 restricting  for simplicity
to first order
inhomogeneous DAs.
These are probably  not
 flexible enough, so that  only partial
improvements are gained and therefore the analysis
ought to be properly extended to  second order DAs.

One should also notice that it is necessary to
analyze various moments in order
to compute all critical parameters of
interest, but not all moments,
(at a given fixed order of HTE)
are expected to be equally reliable. Indeed, for higher
values of $l$, the moment $m^{(l)}(\beta)$
receives a larger contribution from
correlations between distant spins for which
 a smaller number of HTE coefficients is available,
so that it might be slower in reaching the
 asymptotic regime.
In this connection we have also explored
the consequences of a  prescription of least
sensitivity of the results to the choice of $l$.
 As we shall discuss later,  in an estimate
of the  critical exponent $\eta$ which
involves two different moments, it seems  convenient
to use $m^{(0)}(\beta)$ and another moment $m^{(l)}(\beta)$
with small $l$ chosen such that
the estimate of $\eta$ is stationary in $l$.

Let us now specify our way of presenting the results.

Given the first $n+1$ terms of a power
series in $\beta$, we will form all  $ [N/D]$ PAs
with $  N+D \leq n$. We will always take  $N ,D \geq 5$.
The quantities of interest for each approximant,
such as the location of the "critical"
pole, the residue at that pole
or simply the value at $ \beta_{c}$ of the PA,
 will be displayed in a triangular array,
denoted as a Pad\'e table, with N
labelling the columns and D the rows. Whenever
  some entry is followed by an
asterisk ("defective entry")
we will mean that there might be
convergence problems
indicated by  the presence in
the corresponding approximant of
 more than a single pole in the
range $0 \leq \beta \leq 1.2 \beta_{c}$
 or in a narrow  complex strip containing this
segment. A blank is left in the table
 whenever no value in the numerical range of interest
exists for the corresponding approximant.

The indications coming from a  PA analysis using
 $n+1$  coefficients,  will be  summarized
by an estimate, obtained by averaging over a "sample"
including all nondefective entries
of the PA table with $N,D \geq 5 $ and
$n-2 \leq N + D \leq n$
 and qualified by a conventional "error",
defined as twice the standard deviation of the mean value.
Whenever, for sufficiently large $N+D$,
the entries of the PA table are not too scattered
and the successive averages show no residual trend, these
are sensible definitions which
 may be slightly refined
by excluding from our sample occasional
entries differing from the mean  more than five standard
deviations and then recomputing the
mean value and the standard deviation
on the smaller sample.
Sometimes we shall find suggestive to visualize by a
histogram the spread of the entries  of a Pad\'e table.

Let us now evaluate $\epsilon$ by a Pad\'e technique.

Computing the PAs to $ D{\rm ln}[{\rm ln}(\chi)/\beta] $,
the logarithmic derivative of ${\rm ln}(\chi)/\beta$
should  enable us to discriminate between the
structures (\ref{eq:asymom}) and (\ref{eq:asymompow})
of the critical singularity,
since the residues at the critical poles
have either to approach  $ \sigma $, if (\ref{eq:asymom})
holds, or to vanish, if (\ref{eq:asymompow})  holds.

 Table VIII is the Pad\'e table for the
location of the critical pole of
the approximants to
$ D{\rm ln}[ {\rm ln}(\chi)/\beta] $ and Table IX
is the Pad\'e table for the residues.
{}From this (unbiased) analysis we get  the estimates
 $\beta_{c} = 1.118 \pm 0.003$ and
 $\sigma = 0.52 \pm 0.03$.

By a similar argument it should be
convenient to study the
residue at the critical pole for the  PAs
to  $ D{\rm ln}[D{\rm ln}(\chi)]$, the double logarithmic
derivative of $\chi(\beta)$.
 The residue  should  tend either to  1,  if (\ref{eq:asymompow})
holds, or to
$ 1 + \sigma $, if (\ref{eq:asymom}) holds.

 In this case, however, the convergence  is less good,
probably due to subdominant
singularities stronger than in the
previous case, but again the KT structure
is  clearly favoured. We find $\beta_c=1.114 \pm 0.0035$
and $\sigma=0.4 \pm 0.02 $.
We can try to reduce the influence of
the confluent singularities
by a simple modification of
this analysis, namely by
computing PAs to
$\tau D{\rm ln}(\tau D{\rm ln}(\chi))$
at $\beta= \beta_{c}$.
Of course, this is a biased test since
a previous knowledge of
$\beta_{c}$ is required.
If we take
$\beta_{c}=1.118$, we get $\sigma=0.48\pm0.03$,
in good agreement with the results obtained
from the study of $ D{\rm ln}[ {\rm ln}(\chi)/\beta] $.
Repeating these tests on higher order moments we get
 consistent results although with slightly higher
central values for $\beta_{c}$. For instance, a study of
$ D{\rm ln}[{\rm ln}(1+m^{(1)})/\beta] $ yelds
$\beta_{c}=1.127 \pm 0.005$, and $\sigma=0.55 \pm 0.03$
however the successive averages show a residual decreasing trend.

Another simple  (biased) test of the singularity
structure (\ref{eq:asymom}) is  performed
by  computing  the quantity
\begin{equation}
T(\chi) = \frac {D{\rm ln}(D\chi(\beta))} {D{\rm ln}(\chi(\beta))}
\label{eq:t} \end{equation}
or  analogous quantities formed
from other correlation moments.
If  $\chi(\beta)$  has the KT singularity structure
(\ref{eq:asymom}),
then, as $ \beta \uparrow \beta_{c}$,
 we find $T(\chi) =  1+O(\tau^{\sigma})$.
If, on the contrary, $\chi(\beta)$
has a power-law singularity, then
$T(\chi) = 1+\frac {1} {\gamma} + O(\tau)$.
We can therefore distinguish the two cases
by evaluating PAs of  $T(\chi)$ at the critical inverse
temperature $\beta_{c}$
 and thus obtaining  some "effective value" for $\gamma$.
 We have taken $\beta_{c}=1.118$.
The histogram  in Fig.5 shows the
distribution of the values
of  $\frac{1} {T-1}$ in the PA table.
Analogously  computing  $T(m^{(1)})$
 we can obtain an
"effective value" for $\gamma+\nu$.
The results of this computation are also
reported in  Fig.5  as a hatched histogram.
The data suggest that
$ \gamma > 4.5 $ and $ \gamma + \nu > 6.5 $,
 showing complete consistency with
the indications from the ratio tests.
These conclusions remain  essentially unmodified
if we evaluate $T(\chi)$ and $T(m^{(1)})$
at the smaller value  $\beta_{c}=1.09$
which, under the assumption of power-law behavior,
seems to be indicated by
ratio extrapolations and by  PAs to the
logarithmic derivative of $\chi(\beta)$ (see below).
Finally, taking the value $\beta_{c}=1.01$, as
indicated by a recent power-law fit
to MC data [\cite{seiler,edwards}],
yelds $ \gamma > 3.5 $ and  $ \gamma + \nu > 6.5$.
Needless to mention, we may also
compute $T({\rm ln}(\chi))$ or
$T(D{\rm ln}(\chi))$ and check that ${\rm ln}(\chi)$  (respectively
$D{\rm ln}(\chi)$) exhibit the  power
singularities expected from (\ref{eq:asymom}).

Another biased computation which
gives fairly good estimates for some critical
parameters is the following.
Let us fix a value for the constant
$\sigma'$ and compute PAs
for the quantity

\begin{equation}
 [ {\rm ln}(\chi)/\beta]^{1/\sigma'} =
(\frac{ (2-\eta)b}
{\beta_{c}\tau^{\sigma}})^{1/\sigma'}
[1+ O(\tau^{\sigma}{\rm ln}(\tau))]
\label{eq:Sbiased} \end{equation}
under the assumption (\ref{eq:asymom}).

 We have  varied
$\sigma'$ in small steps from 0.49 to 0.51.
The results are summarized in Fig.6
showing how the estimate of
$\beta_{c}$  depends on $\sigma'$.
By the same procedure
we may get the quantity
$ (2-\eta)  b $ as a function of $\sigma'$
from the residue at the critical pole of the PAs.
For the particular value
 $ \sigma' = 1/2 $, the PA table for the
position  of the critical singularity
is reported as Table X. The distribution of the values
of $\beta_c$ in the PA table
is displayed in the histogram
of Fig.7. The values of $b$ computed
from the residues at the
poles (assuming moreover $\eta=1/4$)
are reported in Table XI. From
this analysis final estimates for
$ \beta_{c}$ and $b$ are
$ \beta_{c} =1.1151 + 0.14(\sigma'-0.5) \pm 0.0002 $
and $ b = 1.672 - 3.4(\sigma'-0.5) \pm 0.004$.

If we assume a power-law singularity
like in (\ref{eq:asymompow}), from a study
 of the PAs to the logarithmic
derivative of the susceptibility
$D{\rm ln}(\chi)$, we should be able to estimate $\beta_{c}$,
and from their residues, the critical exponent $\gamma$.
As we have already discussed
[\cite{bcm}], both the PA tables
for the poles and for
the residues (which we shall not report)
contain many "defective entries"
or blanks and do not show  a
 good
convergence. These features of the
approximants   suggest that the critical
singularity is not a power.
If we insist  in producing anyway  some estimate
of the critical parameters,
then, by averaging over all relevant entries
 of the PA tables
for the poles and residues of
the approximants to $D{\rm ln}(\chi)$
 with $ 18 \leq N+D \leq 19 $, we
get  $\beta_{c} = 1.08 \pm 0.02$
and $\gamma = 4.1 \pm 0.6$. Analogously
by studying the PAs of $ D{\rm ln}(m^{(2)}/m^{(1)}) $,  we get the
estimates $\beta_{c}=1.07 \pm 0.01 $ and $\nu = 2.2 \pm 0.1$.
These estimates are  consistent with those obtained from
ratio tests and from a study of  $T(\chi)$ and $T(m^{(1)})$
 but not with those obtained from
recent power-law fits to MC data[\cite{seiler,edwards}].
A histogram showing the broad distribution of
the values of $\beta_{c}$ in the PA table for $D{\rm ln}(\chi)$
is reported in Fig.7.
Notice the  contrast with the very narrow distribution of
the values of $\beta_{c}$ in the PA
table for $({\rm ln}(\chi)/\beta)^{2}$
 displayed (as a hatched histogram)  in the same figure.

The critical index  $\eta$
governing the large distance behavior
of spin-spin correlation
functions may be estimated
observing that,  by eq.
(\ref{eq:corleng}) and (\ref{eq:asymom}),
\begin{equation}
 2-\eta = d{\rm ln} (\chi) /d{\rm ln}(m^{(2)}/m^{(1)})
\label{eq:eta} \end{equation}
at  $ \beta = \beta_{c} $.
 Taking $ \beta_{c} = 1.118 $, yelds
 $ \eta = 0.293 \pm 0.015 $,
 which is not too far from the value predicted by KT.
A similar quantity has also been computed by
MC simulations [\cite{gupta,janke}]
in the range  $0.73 <\beta< 0.94$ obtaining
results completely consistent with ours.
The exponent  $\eta$ may also be computed from
PAs to the ratio

\begin{equation}
 L(\beta,l)= \frac {{\rm ln}(\chi)} {{\rm ln}(1+m^{(l)})}=
\frac {2-\eta} {2-\eta+l} +O(\tau^{\sigma}{\rm ln}(\tau))
\label{eq:eta1} \end{equation}
We have taken $\beta_c=1.118$ and have repeated the computation
for closely spaced values of $l$ in the range
$[1/4,7/4]$. The estimates so obtained
have been plotted vs. $l$ in Fig.8.
Since the results should not depend on $l$, it is reasonable to
expect that the stationary value of
$ L(\beta_{c},l)$ with respect to $l$ is the
best value of $\eta$. This gives $\eta= 0.27 \pm 0.02$

Finally let us  discuss briefly the small power correction
$\tau^{-\theta}$ to the
dominant singular behavior in
(\ref{eq:asymom}) which  has always
eluded  detection by any
numerical method, including HT series.
By taking advantage of the three new HT coefficients available
 we can now give a first indication of its existence.
We can isolate this singularity in a doubly biased analysis
 (with respect to $\eta$ and $\beta_{c}$) by forming the quantity

\begin{equation}
 S(\beta) = \chi(\beta)( m^{(1)}(\beta)
/m^{(2)}(\beta))^{2-\eta}= O( \tau^{-\theta})
\label{eq:Sbeta} \end{equation}
Then
$\theta$ is obtained as the residue of
the logarithmic derivative
of $S(\beta)$ at
$\beta=\beta_{c}$ or, equivalently, as
the value at $\beta_{c}$ of
the quantity
\begin{equation}
\theta(\beta,\eta)=\tau[D{\rm ln}(\chi)+(2-\eta)
D{\rm ln}(m^{(1)}/m^{(2)})]
\label{eq:theta} \end{equation}
In Fig.9 we have reported the quantity
$\theta(\beta_{c},\eta)$ vs. $\eta$
for various values of $\beta_{c}$
 in the range $1.112 <\beta_{c}< 1.121$.
It is remarkable how
 precisely correlated are the expected values of $\theta$
and $\eta$.

\section{ Conclusions }
\label{sec:conc}
Let us now compare our results to those
obtained in previous papers and state our conclusions.

The first MC works [\cite{oldmc}]
were suggestive, but inconclusive.
 The size of the systems studied was
too small and therefore
the range of values of $\beta$ explored in
the simulations was still too far
 away from criticality, so that  the  data,
although  compatible with  KT behavior,
 generally could be
 fitted as well  in terms of a conventional
power-law singularity.

The new generation
of  MC studies [\cite{gupta,wolff,edwards,janke}],
 taking decisive advantage both of
 the greater computing power presently available
 and of the newly invented
algorithms with reduced critical
slowing-down, could be performed
on rather large lattices,  up to $512^{2}$ sites
 [\cite{gupta}] (or even $1200^{2}$ sites  in the
case of Ref. [\cite{janke}] devoted, however, to the
Villain model [\cite{villain}]).
 As a consequence, the recent data are
either practically free from finite size
effects[\cite{gupta,wolff}], [\cite{janke}]
or they have   been carefully [\cite{edwards}]
analyzed in terms
of finite size scaling.
As it has been extensively discussed
in Ref.[\cite{edwards}],
 the recent simulations give
 more reliable,
but, perhaps, not yet definitive
indications in favor of the KT description.
 Indeed  the authors of Ref.[\cite{edwards}]  point out that
 a previous fit [\cite{seiler}]
 to power-law behavior,
which produced
 the following estimates  for the critical
 parameters:
$ \beta _{c}=1.01 \pm 0.01$;
$\gamma = 2.17 \pm 0.10$; $\nu = 1.34 \pm 0.04$;
$\eta = 0.386 \pm 0.02 $,
seems  to be still consistent with their
data, provided that
the critical inverse temperature is increased to the
value $\beta_c=1.05$.

The data of Ref.[\cite{edwards}], of course,  may also be fitted
to the KT behavior, and, assuming $\sigma=1/2$,
the estimates  $\beta_c=1.13 \pm0.015$
and $b=2.15 \pm 0.1$ are obtained.

We should also mention
an overrelaxed MC study on a $512^{2}$ lattice[\cite{gupta}],
in which data have been taken
up to $\beta=1.02$.
The conclusions of Ref.[\cite{gupta}], are
not essentially different:
it is  difficult to discriminate between the KT
and the power-law  fits,
(even) if only MC data
for $\beta \geq 0.94 $  ($\xi > 15$) are used.
Moreover unconstrained independent
(four parameter)  best fits to KT
behavior of the data for $\chi$
and $\xi$ require somewhat different
values for $\beta_c$
(1.127 and 1.117 respectively) and for
$\sigma$ (0.57 and 0.47 respectively).
(A safe estimate of the errors in this case
is  supposed to be
given by the differences of these
results rather than by the much smaller
nominal uncertainties in the fit values.)
If, however, $\sigma$ is held fixed at 0.5, the best fit
values for the remaining critical parameters
 are $\beta_{c}=1.118 \pm 0.005$ and $b=1.7 \pm 0.2$,
which agree  with our own estimates.
 In these analyses some difficulties are met
also with the determination
 of the exponent $\eta$ : if its value is extracted
directly from the parameters of  the fits to $\chi$ and $\xi$, it
 turns  out to be much larger than expected.
 It should be noted, however, that  estimates near to the KT value
are obtained either by resorting
to MC renormalization group (as
shown also in Ref.[\cite{biferale}])
or by using a discretized form of eq.(\ref{eq:eta})
(as shown also in Ref.[\cite{janke}]).
It is  interesting to quote also a very recent "verification
 of the KT scenario"[\cite{hasen}] by a method based on matching
the renormalization group flow of the dual of the XY model
 with the flow of the body-centered solid-on-solid model
which has been exactly solved[\cite{baxter}]
to exhibit a KT transition. The method has to assume that
 $\sigma=1/2$ and yields $ \beta_c=1.1197 \pm 0.0005$ and
$b=1.88 \pm0.02$.

 We can summarize the main limitations
of these MC works as follows:
the range of values of $\beta$  covered,
 even in the most extensive
among these studies[\cite{gupta}] is presently
 restricted to $\beta  \leq 1.02 $, ( corresponding to
$ \xi \leq 70.$) and, anyway, the estimates of the critical
parameters have not yet reached
a satisfactory level
of precision. Finally, it should also
be noticed that there are arguments
indicating that the KT critical region has a
very small width [\cite{greif}],
and therefore it might have been explored
only in its extreme periphery by these studies.
Therefore all authors of MC works
agree that further simulations
much closer to $\beta_c$ are still desirable.

 Early HT studies [\cite{htstudies}],
 based on ten term series,
were also inconclusive
 and  unable to  provide  reasonably stable estimates
of the critical parameters.
Better
suited methods of series analysis[\cite{gut}]
were later proposed.
Extensive
calculations on highly
asymmetric lattices [\cite{hamer}] by
hamiltonian strong coupling or
finite lattice techniques always gave  indications
in favor of the KT scenario.

As soon as we  made available
substantially longer HT series [\cite{bcm}],
 it emerged  not only
 that the KT critical behavior is favored by all tests
(this  conclusion was strengthened by the independent
analysis of Ref.[\cite{ferer}] by the four-fit method),
but also  that any series analysis  designed to extract
 power-law
scaling, leads to  estimates of  the critical
parameters  definitely inconsistent with those coming from
the corresponding fits to MC data.
As we have observed above,
 the study of ratio plots, and of
the PAs to the logarithmic derivative of $\chi$ and $\xi$,
   shows clear signs of a  non power-like
nature of the critical singularity,
 and, if forced to produce estimates of
the critical parameters, points to  values
of $\beta_{c}$, $\gamma$ and $\nu$ which are significantly
larger than those  derived
from  power-law fits to MC data
and show a clear trend to increase
with the number of HT coefficients used.
This is precisely what one should
 see when trying to interpret  an infinite
order singularity as a power singularity.
 It is then  reasonable to expect
that also future MC studies
will still appear to be
unable to decide between the two kinds of
fits, but
  power-law fits  will suggest
embarassingly larger and larger values for $\gamma$ and for $\nu$.
Conversely,
 extrapolations of the HT series by ratio, Pad\'e  or
differential approximant techniques,
 in a way consistent with the
KT behavior, show a good stability,
 and agree with the KT fits to the MC
data. They lead to the following values
 of the critical parameters
 $\beta_{c}=1.118 \pm 0.003$, $b=1.67 \pm 0.04$,
$\sigma=0.52 \pm 0.03 $ and $\eta= 0.27 \pm 0.02$,
 and moreover make a first detection of the exponent $\theta$ possible.
 These are  quite acceptable estimates,
although they do not yet reach the
 high level of precision that is usually
expected from so long HT series, probably
because we are not yet using methods of series analysis
entirely adequate to the complicated
nature of the critical singularities.

We conclude that a sufficient extension of the
HT series has been achieved
 to enable us to assert
 that the  critical behavior of the plane rotator model
agrees only
 with the KT predictions [\cite{kt}]
and that it is now  possible,
by various essentially different methods,
to get consistent and fairly accurate
estimates of the critical parameters.

\nonum
\section{Acknowledgments}
\label{ack}
We  thank our friend G.Marchesini for his constant
encouragement and  interest in this project
which we initiated together several years ago.

Our work has been partially supported by MURST.

\figure{The alternate ratios of HTE coefficients
of various moments are plotted vs. $1/n$.
 The alternate ratios $ \bar r_{n}(\chi) $ are represented by
black squares; $\bar r_{n}(m^{(1)}) $ by black triangles.
 We  have also plotted the  linearly extrapolated sequences
$ \bar r_{n}^{(1)}(\chi) $
(empty squares), and  $\bar r_{n}^{(1)}(m^{(1)}) $  (empty
triangles).\label{ratioplot}}
\figure{ Unbiased estimates of the critical exponent $\gamma$ of
the susceptibility under the assumption of a power-law critical
singularity
obtained from the alternate ratios $ \bar r_{n}(\chi) $
(empty circles). Analogous estimates of the exponent $\gamma + \nu$
 as obtained from $\bar r_{n}(m^{(1)}) $ (empty triangles).
\label{gammapower}}
\figure{Ratio plots for alternate HTE coefficients vs.
$1/n^{2/3}$.
 The alternate ratios $ \bar r_{n}(\chi) $ are represented by
black squares; $\bar r_{n}(m^{(1)}) $ by black triangles.
The alternate ratio sequences  have been
 extrapolated in $1/n^{2/3}$ obtaining the sequences
$ \bar s_{n}(\chi) $
(empty squares), $\bar s_{n}^{(1)}(m^{(1)}) $  (empty
triangles).
 A further extrapolation in $1/n$
of the sequences $s_{n}$ gives $ \bar s_{n}^{(1)}(\chi) $
(black stars), $\bar s_{n}^{(1)}(m^{(1)}) $
(crosses).
\label{ratioplott}}
\figure{ The sequences $\epsilon_{n}$ (crosses),
 $\epsilon_{n}'$ (empty triangles),
$\epsilon_{n}''$ (empty circles), as
computed from the quantities $t_{n}$ introduced in (\ref{eq:ratioeps}),
and from the analogous ones $ u_{n}$
and $v_{n}$,  are plotted  versus $\frac {1} {n} $.
 The dashed line indicates the KT prediction for the
 value of $\epsilon$.\label{epsilon}}
\figure{ Distribution of the values in the PA table of the
 quantity $(T(\chi)-1)^{-1}(\beta_c)$ defined by (\ref{eq:t})
(unhatched histogram). The same for the quantity
$(T(m^{(1)})-1)^{-1}(\beta_c)$
(hatched histogram).\label{tisto}}
\figure{  Estimates of the
inverse critical  temperature $\beta_{c}$ (biased in $\sigma$),
obtained from a study  of the "critical pole" of the PAs to
$(\frac {{\rm ln}(\chi)} {\beta} )^{\frac {1} {\sigma}}$,
are plotted versus $\sigma$.\label{logchi}}
\figure{ Histogram of the distribution
of the values of $\beta_{c}$
in the  table of PAs to $D{\rm ln}(\chi)$
($N,D \geq 5$ and $N+D \leq 19$)
 with a resolution of $10^{-3}$.
For contrast the narrow distribution
of the values of $\beta_{c}$ in the
 PA table for $({\rm ln}(\chi)/\beta)^2$
is also reported (hatched histogram
on the right).\label{histogram}}
\figure{ The estimates of the exponent $\eta$
 obtained from the quantity $L(\beta_{c},l)$ of
eq.(\ref{eq:eta1}) are plotted
vs. $l$ for $\beta_{c}=1.118$.
 The continuous line indicates the KT prediction for the
 value of $\eta$.\label{lbeta}}
\figure{ The estimates (biased in $\eta$ and
$\beta_{c}$) of the exponent $\theta$
as computed from the quantity
$\theta(\beta,\eta)$ defined in (\ref{eq:theta})
 are plotted versus $\eta$
for various values of $\beta_{c}$:
$\beta_{c}=1.112$ (squares), $\beta_{c}=1.115$ (triangles),
 $\beta_{c}=1.118$ (triangles),
$\beta_{c}=1.121$ (rhombs).
 The continuous line indicates the KT prediction for the
 value of $\theta$.
 \label{ftheta}}
\narrowtext
\begin{table}
\caption{HTE coefficients of the
nearest neighbor  correlation  $C(0,x)$ with $x=(0,1)$}\label{Cnn}
\begin{tabular}{cc}
order&coefficient\\
\tableline
          1  &0.500000000000000000000000000000\\
          3  &0.187500000000000000000000000000\\
          5  &0.010416666666666666666666666666\\
          7& -0.005045572916666666666666666666\\
          9& -0.011897786458333333333333333333\\
         11& -0.009914482964409722222222222222\\
         13& -0.006428721594432043650793650793\\
         15& -0.003556433509266565716455853174\\
         17& -0.001900080568517583644836294214\\
         19& -0.000804827256075995542618566322\\
\end{tabular}
\end{table}
\narrowtext
\begin{table}
\caption{HTE coefficients of the
next nearest neighbor  correlation  $C(0,x)$ with $x=(0,2)$}\label{C02}
\begin{tabular}{cc}
order&coefficient\\
\tableline
          2 &0.25000000000000000000000000000000\\
          4 &0.31250000000000000000000000000000\\
          6 &0.04557291666666666666666666666666\\
          8& -0.0239257812500000000000000000000\\
         10& -0.0236504448784722222222222222222\\
         12& -0.0180172390407986111111111111111\\
         14& -0.0109599196721637059771825396825\\
         16& -0.00643771678156743394424465388007\\
         18& -0.00331282301929834284736622835219\\
         20& -0.00150452549535034334988961284262\\
\end{tabular}
\end{table}
\narrowtext
\begin{table}
\caption{HTE coefficients of the
next nearest neighbor  correlation:  $C(0,x)$ with $x=(1,1)$}\label{C11}
\begin{tabular}{cc}
order&coefficient\\
\tableline
          2 &0.500000000000000000000000000000\\
          4 &0.125000000000000000000000000000\\
          6 &0.005208333333333333333333333333\\
          8& -0.012695312500000000000000000000\\
         10& -0.016959635416666666666666666666\\
         12& -0.012420654296875000000000000000\\
         14& -0.007721207633851066468253968253\\
         16& -0.004255058809562965675636574074\\
         18& -0.002158290178150189710135275818\\
         20& -0.000850845285942630162314763144\\
\end{tabular}
\end{table}
\narrowtext
\begin{table}
\caption{HTE coefficients of the
susceptibility $m^{(0)}$}\label{mom0}
\begin{tabular}{cc}
order&coefficient\\
\tableline
 0    &1.000000000000000000000000\\
 1    &2.000000000000000000000000\\
 2    &3.000000000000000000000000\\
 3    &4.250000000000000000000000\\
 4    &5.500000000000000000000000\\
 5    &6.854166666666666666666667\\
 6    &8.265625000000000000000000\\
 7    &9.722005208333333333333333\\
 8    &11.205078125000000000000000\\
 9    &12.675553385416666666666667\\
10    &14.152012803819444444444444\\
11    &15.601900227864583333333333\\
12    &17.019300672743055555555556\\
13    &18.392466299874441964285714\\
14    &19.714506515624031187996032\\
15    &20.971455838629808375444362\\
16    &22.163650634196279751140184\\
17    &23.280944825182959960064373\\
18    &24.320568285114725921379686\\
19    &25.279185763955802490448171\\
20    &26.153731926768238512226443\\
\end{tabular}
\end{table}

\narrowtext
\begin{table}
\caption{HTE coefficients of the
first correlation moment $m^{(1)}$}\label{mom1}
\begin{tabular}{cc}
order&coefficient\\
\tableline
 0   &0.000000000000000000000000\\
 1   &2.000000000000000000000000\\
 2   &4.828427124746190097603377\\
 3   &8.958203932499369089227521\\
 4   &14.774302788642591334803698\\
 5   &22.405537350785873843406389\\
 6   &32.018311604539175300778647\\
 7   &43.776633965886723037276578\\
 8   &57.804795726728279225339368\\
 9   &74.171223440617174956388203\\
10   &92.948162559177956624043786\\
11   &114.170878181118055761226873\\
12   &137.820848184699262000940865\\
13   &163.889685030219313242632769\\
14   &192.312469972536223823702791\\
15   &223.008032736766111405219191\\
16   &255.881465307579934186740726\\
17   &290.805253343564738390269085\\
18   &327.634680720388208255919862\\
19   &366.217176852355184083122429\\
20   &406.375299064909623959197097\\
\end{tabular}
\end{table}
\narrowtext
\begin{table}
\caption{HTE coefficients of the
second correlation moment $m^{(2)}$}\label{mom2}
\begin{tabular}{cc}
order&coefficient\\
\tableline
 0   &0.000000000000000000000000000\\
 1   &2.000000000000000000000000000\\
 2   &8.000000000000000000000000000\\
 3   &20.25000000000000000000000000\\
 4   &42.00000000000000000000000000\\
 5   &76.85416666666666666666666666\\
 6   &129.0208333333333333333333333\\
 7   &203.2220052083333333333333333\\
 8   &304.6718750000000000000000000\\
 9   &438.9568033854166666666666666\\
10   &612.0054470486111111111111111\\
11   &830.0374037000868055555555556\\
12   &1099.397710503472222222222222\\
13   &1426.589506772964719742063492\\
14   &1818.089718954903738839285714\\
15   &2280.298322941197289360894097\\
16   &2819.491738309136984419780644\\
17   &3441.674843107074259683802248\\
18   &4152.534972385628279951911521\\
19   &4957.398607418558360103903951\\
20   &5861.100409957330553479786132\\
\end{tabular}
\end{table}
\begin{table}
\caption{HTE coefficients of the
expectation of the squared
vorticity: $<v(\beta)^{2}>$}\label{vort}
\begin{tabular}{cc}
order&coefficient\\
\tableline
        0  &0.33333333333333333333333333333333\\
        1& -0.20264236728467554288775892641946\\
        2& -0.13931662750821443573533426191338\\
        3& -0.00093815910779942380966555058527\\
        4& -0.01391846988836801417621000438623\\
        5  &0.02145616787492672223895597466054\\
        6  &0.00395517130108460209757177042318\\
        7  &0.00626571732016345480145945834089\\
        8  &0.00498164199972076876178153427788\\
        9 &0.00188487896531662245169702329601\\
        10 &0.00506374324797059743179508168123\\
        11& -0.00016501166685598670821344313122\\
        12 &0.00321456708015923485554342658920\\
        13& -0.00038482795295650133767225429558\\
        14 &0.00167133027095100840999836256202\\
        15& -0.00028727125412299355443537746859\\
\end{tabular}
\end{table}
\widetext
\begin{table}
\caption{The location of the critical poles of the PAs to
$D{\rm ln}({\rm ln}(\chi)/\beta)$. The degree N of the PA
numerator labels the columns and the degree D of the denominator
labels the rows. Asterisks indicate the "defective entries",
 blanks indicate the lack of an acceptable entry.}\label{cpdll}
\begin{tabular}{cccccccccc}
&\multicolumn{1}{c}{N}&&&&&&&&\\
D&5&6&7&8&9&10&11&12&13\\
\tableline
5&  1.1408   &1.1277  &1.1129   &
  &1.1312   &1.1319*   &1.1029  &1.1087  &1.1133\\
6&  1.1476*   &   &1.1193   &1.1261   &1.1169   &1.1231
  &1.1094   &1.0996*&\\
7&      &1.1431   &1.1225   &1.1223   &1.1219   &1.1202   &1.1174&&\\
 8&  1.1142   &1.1271   &1.1223*   &1.1226*   &1.1225*   &1.1116&&&\\
 9&  1.1208   &1.1242   &1.1218   &1.1225*   &1.1225*&&&&\\
 10&  1.1265   &1.1233   &1.1375*   &1.1029*&&&&&\\
 11&  1.1185   &1.1218   &1.1068*&&&&&&\\
 12&  1.1236   &1.1208&&&&&&&\\
  13& 1.1611* &&&&&&&&\\
\end{tabular}
\end{table}
\begin{table}
\caption{The residues at the critical poles of the PAs to
$D{\rm ln}({\rm ln}(\chi)/\beta)$. Same conventions as in the previous
table.}\label{resdll}
\begin{tabular}{cccccccccc}
&\multicolumn{1}{c}{N}&&&&&&&&\\
D&5&6&7&8&9&10&11&12&13\\
\tableline
5&  0.611&  0.563&  0.504&  &  0.596&  0.600*&
0.420&  0.458&  0.488\\
6&  0.630*&&  0.531&  0.564&  0.516&
0.549&  0.463&  0.400*&\\
7&&  0.620&  0.545&  0.545&  0.542&
0.533&  0.516&&\\
8&  0.512&  0.567&  0.545*&  0.546*&
0.545*&  0.470&&&\\
9&  0.540&  0.554&  0.542&  0.545*&  0.546*&&&&\\
10&  0.566&  0.550&  0.549*&  0.364*&&&&&\\
11&  0.526&  0.543&  0.408*&&&&&&\\
12&  0.553&  0.537&&&&&&&\\
13&  0.740*&&&&&&&&\\
\end{tabular}
\end{table}
\begin{table}
\caption{The location of the critical poles of the PAs to
$({\rm ln}(\chi)/\beta)^2$.  Same conventions as in the previous
table.}\label{cpll}
\begin{tabular}{ccccccccccc}
&\multicolumn{1}{c}{N}&&&&&&&&&\\
D&5&6&7&8&9&10&11&12&13&14\\
\tableline
5&   &   1.1088&   1.1105  & 1.1106 &&  & 1.1167
& 1.1154  & 1.1149  & 1.1149\\
6&    1.1152  & 1.1132 & 1.1119  & 1.1079* &&  1.0811*
& 1.1151  & 1.1140  & 1.1149&\\
7&    1.1136  & 1.0901*  & 1.1127  & 1.0944*  & 1.1091
& 1.1128  & 1.1145  & 1.1148&&\\
8&    1.1128  & 1.1130  & 1.1136
& 1.1153  & 1.1160  & 1.1155  & 1.1154&&&\\
9&    1.1130  & 1.1127*  & 1.1059*
& 1.1164  & 1.1157  & 1.1154&&&&\\
10&    1.1142  & 1.1189*  & 1.1148
& 1.1147  & 1.1155&&&&&\\
11&    1.1155  & 1.1155  & 1.1147  & 1.1148&&&&&&\\
12&    1.1155  & 1.1155  & 1.1152&&&&&&&\\
13&    1.1173*  & 1.1151&&&&&&&&\\
14&    1.1157&&&&&&&&&\\
\end{tabular}
\end{table}

\begin{table}
\caption{The residues at the critical poles of the PAs to
$({\rm ln}(\chi)/\beta)^2$.  Same conventions as in the previous
table.}\label{crll}
\begin{tabular}{ccccccccccc}
&\multicolumn{1}{c}{N}&&&&&&&&&\\
D&5&6&7&8&9&10&11&12&13&14\\
\tableline
5& & 1.609& 1.623& 1.623&&& 1.692& 1.675& 1.668&1.668\\
6& 1.670 & 1.649 & 1.636&1.607*&&2.255*& 1.671& 1.655 & 1.669&\\
7& 1.653  &1.306*&1.644& 1.676*& 1.613& 1.643 &1.663&1.667&&\\
8& 1.646  & 1.648&1.654 & 1.674& 1.685& 1.677& 1.676&&&\\
9& 1.648 & 1.645*&1.636*&  1.692& 1.680& 1.675&&&&\\
10& 1.660 & 1.738*& 1.667& 1.666& 1.677&&&&&\\
11& 1.676 &1.677& 1.665& 1.667&&&&&&\\
12& 1.677 & 1.676&1.673&&&&&&&\\
13& 1.698* & 1.671&&&&&&&&\\
14& 1.680 &&&&&&&&&\\
\end{tabular}
\end{table}
\end{document}